\documentstyle[aps,preprint,psfrag,epsfig]{revtex}
\newcommand{\be}{\begin{equation}}
\newcommand{\ee}{\end{equation}}
\newcommand{\ba}{\begin{eqnarray}}
\newcommand{\ea}{\end{eqnarray}}
\newcommand{\no}{\nonumber \\}
\begin{document}
\begin{titlepage}
\pagestyle{empty}
\vspace{1.0in}
\begin{flushright}
October 1999
\end{flushright}
\vspace{.1in}
\begin{center}
\begin{large}
{\bf The cosmology with the Dp-brane gas}
\end{large}
\vskip 0.5in
Chanyong Park\footnote{ E-mail : chanyong@hepth.hanyang.ac.kr}, 
 Sang-Jin Sink\footnote{ E-mail : sjs@hepth.hanyang.ac.kr}  
 and Sunggeun Lee\footnote{ E-mail : sglee@hepth.hanyang.ac.kr}\\
\vskip 0.2in
{\small {\it Department of Physics, Hanyang University\\
Seoul, Korea}}
\end{center}
\vspace{1 cm}

\begin{abstract}
We study the effect of the Dp-brane gas in string
cosmology. When one kind of Dp-brane gas dominates
, we find that the cosmology is equivalent to that of the
Brans-Dicke theory with the perfect fluid type matter. 
We obtain $\gamma$, the equation of state parameter, in terms of $p$ and 
the space-time dimension. 
\end{abstract}  

\vspace{.3in}\noindent PACS numbers: 98.80.Cq, 04.50.+h
\end{titlepage}

Recent developments of the string theory suggest that in a 
regime of Planck length, quantum fluctuation is very large  
so that string coupling becomes large and consequently 
the fundamental string degrees of freedom
are not a weakly coupled {\it good} ones \cite{witten}.
Instead, solitonic degrees of freedom like p-branes or 
Dp-branes \cite{pol} are more important. 
Therefore it is a very interesting question 
to ask what is the effect of these new degrees of freedom to 
the space time structure especially whether including these degrees 
of freedom resolves the initial singularity, which is a problem in 
standard general relativity.

What should be the starting point for the investigation of  the
p-brane cosmology? It should be a generalization of general
relativity. The Brans-Dicke theory is a generic deformation of the
general relativity allowing variable gravity coupling. 
In fact the low energy theory of the fundamental string \cite{vene} 
contains the Brans-Dicke (BD) theory with a fine tuned deformation parameter
($\omega$=-1). Moreover Duff and et al. \cite{du} found that the
natural metric that couples to the p-brane is the  Einstein metric
multiplied by certain power of dilaton field.  In terms of this new
metric, the action that gives the p-brane solution  becomes
BD action with definite deformation  parameter $\omega$
depending on p.
In our previous papers \cite{cyong,sgl}, we have studied this
kind of the BD model and found exact solutions depending on two
arbitrary parameters. Furthermore, according to
the range of parameters involved, we also have classified all possible
behaviors of the BD cosmology and found solutions resolving the
initial singularity
problem for some regions \cite{ra}. However, it remains an open questions
how to determine the equation of state parameter $\gamma$. In this paper,
we will determine $\gamma$ from the relation between this BD theory and
the string theory with Dp-brane. 

We start with reviewing the BD theory in which the perfect
fluid type matter is included. Let's assume that our universe is a $D$
dimensional homogeneous isotropic one. The general BD action is given by 
\be
S = \int d^D x \sqrt{-g} e^{-\phi} \left[ {\cal R} - \omega \nabla_{\mu} \phi
\nabla^{\mu} \phi \right] + S_m ,
\ee
where $\phi$ is the dilaton field and $S_m$ is the
matter part of the action.
The equations of motion of the BD theory become
\cite{Weinberg,Veneziano}
\ba
{\cal R}_{\mu\nu} - \frac{g_{\mu\nu}}{2} {\cal R} &=&
       \frac{e^{\phi}}{2} T_{\mu\nu} + \omega  \{ \nabla_{\mu} \phi
       \nabla_{\nu} \phi - \frac{g_{\mu\nu}}{2} (\nabla \phi)^2
       \} \no 
&& + \{ -\nabla_{\mu} \nabla_{\nu} \phi +
       \nabla_{\mu} \phi \nabla_{\nu} \phi + g_{\mu\nu} {\cal D}^2
       \phi - g_{\mu\nu} (\nabla \phi)^2 \} , \no
0 &=& {\cal R} - 2\omega {\nabla}^2 \phi + \omega (\nabla \phi)^2 .\label{bdeq} 
\ea
${\cal R}$ is the curvature scalar and cosmological metric
is given by
\be
{ds_D}^2 = -{\cal N} dt^2 + e^{2\alpha(t)} \delta_{ij} dx^i dx^j \;\; 
( i,j = 1, 2, \cdots, D-1) , \label{met}
\ee
where $e^{\alpha(t)}$ is the scale factor and ${\cal N}$ is
a (constant) lapse  function. Now, we assume that all variables are
the functions of time $t$ only, then the curvature scalar \cite{Lu} in
$D$ dimension is given by
\be
{\cal R} = g^{00} {\cal R}_{00} + g^{ij} {\cal R}_{ij},
\ee
where
\ba
g^{00} {\cal R}_{00} &=& \frac{D-1}{\cal N} [ \ddot{\alpha} 
                             + \dot{\alpha}^2 ], \nonumber \\
g^{ij} {\cal R}_{ij} &=& \frac{D-1}{\cal N} [ \ddot{\alpha} 
                             + (D-1) \dot{\alpha}^2 ]
\ea 
whit $\dot{\alpha} = \nabla_t \alpha$. 
With the equation of state $p=\gamma \rho$,  
the energy-momentum tensor of the perfect-fluid type matter is given by 
\begin{equation} 
T_{\mu\nu} = p g_{\mu\nu} + (p + \rho) U_{\mu} U_{\nu}
\end{equation}            
where $U_{\mu}$ is the fluid velocity. Under the hydrostatic equilibrium
condition, the energy-momentum conservation is
\begin{equation}
\dot{\rho} + (D-1) (p + \rho) \dot{\alpha} = 0 . \label{ceqs}
\end{equation} 
Using $p=\gamma \rho$, we get the solution 
\be 
\rho = \rho_0 e^{-(D-1)(1+\gamma) \alpha}.  
\ee
In our previous papers, the parameters $\gamma$ and $\omega$ were
considered as free parameters. In this paper, however, $\gamma$ will be
related to the dimension of the world volume of p-brane.

If we consider only the time dependence, the equations of motion (\ref{bdeq})
and the energy-momentum conservation (\ref{ceqs}) follow from the action
\ba
S &=& \int dt  e^{(D-1)\alpha-\phi} \left[ \frac{1}{\sqrt{\cal N}}
      \big{\{} -(D-2)(D-1)\dot{\alpha}^2 + 2(D-1) \dot{\alpha}
      \dot{\phi} + \omega\dot{\phi}^2 \big{\}} \right. \no     
  && \left. \hspace{2.5cm} - {\sqrt{\cal N}} \rho_0
      e^{-(D-1)(1+\gamma) \alpha + \phi} \frac{}{} \right]. 
\ea
Now, we introduce a new time variable $\tau$ by 
\be
dt = e^{(D-1)\alpha - \phi} d\tau . \label{deft}
\ee
Then the action can be written as     
\ba
S = \int d\tau && \left[ \frac{1}{\sqrt{\cal N}} \left( -(D-2)(D-1) 
{\alpha^{\prime}}^2 + 2 (D-1) \alpha^{\prime} \phi^{\prime} + \omega
{\phi^{\prime}}^2 \right) \right. \no
&& \left. - \sqrt{\cal N} \rho_0 e^{(D-1)(1-\gamma)\alpha - \phi}
\frac{}{} \right] , \label{bda}
\ea
where `prime' means the derivative with respect to the new time
variable $\tau$ and $\rho_0$ is considered as a positive real
constant. Note that the variation over the constant lapse function
gives a constraint equation. So far we discussed BD theory.  

Now we consider the string cosmology by the D-dimensional effective
action with n-form field strength that is coming from the appropriate
compactification of 10-dimensional low energy effective string theory.
In Einstein frame, the action reads
\be
S = \int d^D x \sqrt{-g^E} \left[ R-\frac{1}{2} \nabla_{\mu} \phi
\nabla^{\mu} \phi  - \frac{e^{-\chi(n) \phi}}{2 n!} {H_n}^2 \right],
\ee 
where $\chi$ is given by
\be
\chi^2 (n) =  4 - \frac{2(n-1)(\tilde{n} -1)}{n+\tilde{n}-2},
\ee
with the dual dimension $\tilde{n} = D -n$.
From this action, we can obtain an action in other frame by a conformal mapping.
The p-brane frame is defined by the conformal mapping
\be
{g^p}_{\mu\nu} = e^{\frac{\chi(n) \phi}{n-1}} {g^E}_{\mu\nu},
\ee
where ${g^E}_{\mu\nu}$ (${g^p}_{\mu\nu}$) means the metric in Einstein
(p-brane) frame. In this p-brane frame \cite{du}, the action is written as
\be   
S = \int d^D x \sqrt{-g^p} e^{-\frac{(D-2) a(n) \phi}{2(n-1)}}
     \left[ R-\omega \nabla_{\mu} \phi
     \nabla^{\mu} \phi  - \frac{1}{2 n!} {H_n}^2 \right],
\ee
where $\omega$ becomes \cite{du}
\be
\omega = - \frac{(D-1)(n-3)-(n-1)^2}{(D-2)(n-3)-(n-1)^2} . 
\ee
In 4-dimension, the BD parameter $\omega$ is given by $\omega =
-\frac{4}{3}$ for the $0$-brane ($p=0$) and $\omega = - \frac{3}{2}$
for the instanton ($p=-1$), etc. 
Hence, the BD parameter $\omega$ can be varied according to the
p-branes included. However, it is not clear whether the
matters come from NSp- or Dp-branes in p-brane frame.
 
In string frame, by an appropriate conformal mapping, the action is
given by
\be
S = \int d^D x \sqrt{-g} \left[ e^{-\phi} \{R + \nabla_{\mu} \phi
\nabla^{\mu} \phi \} - \frac{e^{m\phi}}{2 n!} H_{\mu_1 \cdots \mu_n}
H^{\mu_1 \cdots \mu_n} \right] . \label{sact}
\ee
Notice that the BD parameter $\omega$ is fixed as $-1$. 
For $m=-1$, an n-form field strength comes from the
compactification of NS-NS 3-form in 10-dimensional theory. $m=0$ for
n-form coming from R-R sector. The dual form in NS-NS sector is
defined by \cite{du}
\be
*H = e^{-\phi} H,
\ee
for solitonic NSp-brane, and $m=1$ for this case.

Under the ansatz,
\be
H_{\mu_1 , \cdots , \mu_n} = \sqrt{\cal N} \sqrt{n}
\epsilon_{\mu_1 ,\cdots,\mu_{n-1}} \nabla_0 A(t)
\ee
where $\mu_1 ,\cdots,\mu_{n-1} \ne 0$, then the Bianchi identity 
\be
\nabla_{[\mu} H_{\mu_1 , \cdots , \mu_n ]}= 0
\ee
is always satisfied, since we assume that $A(t)$ is a function of $t$ only.  
$\sqrt{\cal N}$ should be used for obtaining the correct constraint
equation. 
In terms of $A(t)$, $\phi(t)$ and $\alpha(t)$, the action (\ref{sact}) for
the cosmology becomes
\ba
S &=& \int dt  e^{(D-1)\alpha-\phi} \left[ \frac{1}{\sqrt{\cal N}}
      \big{\{} -(D-2)(D-1)\dot{\alpha}^2 + 2(D-1) \dot{\alpha}
      \dot{\phi} - \dot{\phi}^2 \big{\}} \right. \no     
  && \left. \hspace{2.5cm} + \frac{\sqrt{\cal N}}{2} \dot{A}^2
      e^{-2(n-1)\alpha + (m+1)\phi} \frac{}{} \right]. 
\ea
Using the new time variable $\tau$ defined in (\ref{deft}), the action is given by 
\ba
S = \int d\tau && \left[ \frac{1}{\sqrt{\cal N}} \left( -(D-2)(D-1) 
    {\alpha^{\prime}}^2 + 2 (D-1) \alpha^{\prime} \phi^{\prime} -
    {\phi^{\prime}}^2 \right) \right. \no
&&  \left.  + \frac{\sqrt{\cal N}}{2} {A^{\prime}}^2
      e^{-2(n-1)\alpha + (m+1)\phi} \frac{}{} \right] , 
\ea
From this, we can obtain the equations of motion
\be
2(D-1)(D-2) \alpha^{\prime \prime} - 2(D-1) \phi^{\prime \prime} 
- (n-1) (A^{\prime})^2 e^{-2(n-1)\alpha +(m+1) \phi}
= 0  , \label{eq1}
\ee
\be
2(D-1) \alpha^{\prime \prime} - 2 \phi^{\prime \prime} 
- \frac{(m+1)}{2} (A^{\prime})^2 e^{-2(n-1)\alpha
+(m+1) \phi} = 0  , \label{eq2}
\ee
\be
\nabla_{\tau} \left[ A^{\prime} e^{-2(n-1)\alpha 
+ (m+1) \phi} \right] = 0,  \label{str}
\ee
where the lapse function ${\cal N}$ is set to $1$ after calculation. 
The solution of Eq. (\ref{str}) is given by
\be  
A^{\prime} = \sqrt{2} q e^{2(n-1)\alpha -(m+1)\phi} ,
\ee
with a constant $q$. Substituting this into
Eq. (\ref{eq1}) and (\ref{eq2}), the resulting equations of
motion are written as
\be
2(D-1)(D-2) \alpha^{\prime \prime} - 2(D-1) \phi^{\prime \prime} 
- 2 (n-1) q^2 e^{2(n-1)\alpha -(m+1) \phi} = 0  ,
\ee
\be
2(D-1) \alpha^{\prime \prime} - 2 \phi^{\prime \prime} 
- (m+1) q^2 e^{2(n-1)\alpha -(m+1) \phi} = 0 .
\ee
These equations of motion can be derived from the following action
\ba 
S = \int d\tau && \left[  \frac{1}{\sqrt{\cal N}} \left( -(D-2)(D-1) 
{\alpha^{\prime}}^2 + 2 (D-1) \alpha^{\prime} \phi^{\prime} - 
{\phi^{\prime}}^2 \right) \right. \no
&& \left. - \sqrt{\cal N} q^2 e^{2(n-1)\alpha -(m+1) \phi}
\frac{}{} \right] .
\ea
Comparing this action with that of the BD theory given by
Eq. (\ref{bda}), the action of the BD theory is equivalent to that of the
string theory with the Dp-brane gas ($m=0$) if we set $\rho_0 =q^2$  
and $2(n-1) = (D-1)(1-\gamma)$. From these, the BD parameter
$\gamma$ is related to $p$ the spatial dimension of the world volume
of p-brane;
\be
\gamma = \frac{D-2p-3}{D-1}.
\ee
This is the main result of our paper.
Note that in string frame, the behaviors of the cosmology depend only
on $p$ ($=n-2$). In the four dimensional space time, $\gamma$ of the
instanton gas ($p=-1$) is fixed to $1$ and for the particle ($p=0$)
$\gamma = 1/3$ is consistent with the known value of $\gamma$ for the
radiation dominant era. In our previous work \cite{cyong}, 
the behavior of the scale factor for general $\gamma$ and $\omega$ was
classified. With the D-particle gas ($p=0$ and $\gamma = 1/3$),
the string cosmology has two phases as Fig. 1. 
For the D-string gas, $\gamma = -1/3$ which is consistent with
the value of cosmic string gas \cite{sgam}, the behavior
of scale factor has two phases as shown in Fig. 2.
For D2-brane or D3-brane cases ($p=2$ or $p=3$), the scale
factor behaves as Fig. 3. 
 
In this paper, we have shown that the string theory
with Dp-brane gas can be described by the BD theory with the perfect
fluid type matter and the parameter $\gamma$ in BD theory is
determined by the dimension of the brane. 
In string theory with the NS- ($m=-1$) or the dual NS-brane ($m=1$)
gas, the matter couples to the dilaton, and the
energy-momentum of the perfect fluid is not conserved due to this
coupling.
Hence we can not consider the NS type brane gas as a
perfect fluid. To describe the NS brane gas as the perfect fluid, 
we have to study the
string theory in a frame where the dilaton does not couple with
the the brane. Details will be discussed in later work \cite{cyong2}. 
 
\vskip 1cm

\noindent{\bf \Large Acknowledgement}

\noindent This work has been supported by the research grant of
Hanyang University 1999.

\newpage
\begin{figure}
  \unitlength 1mm
   \begin{center}
      \begin{picture}(70,100)
      %\graphpaper[2](-60,0)(190,80)
      \put(-50,10){\epsfig{file=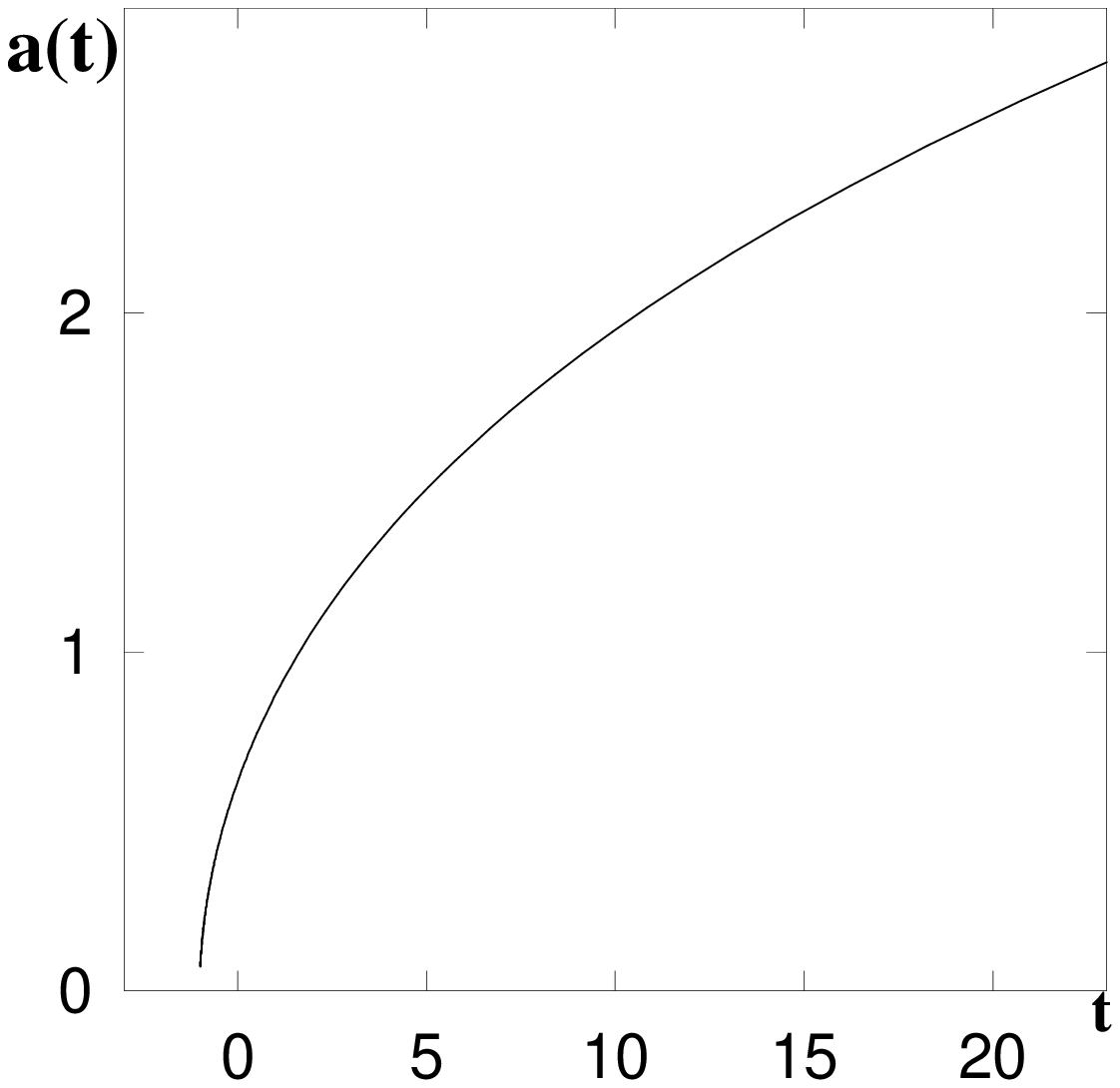,width=10cm,height=12cm}}
      \put(30,10){\epsfig{file=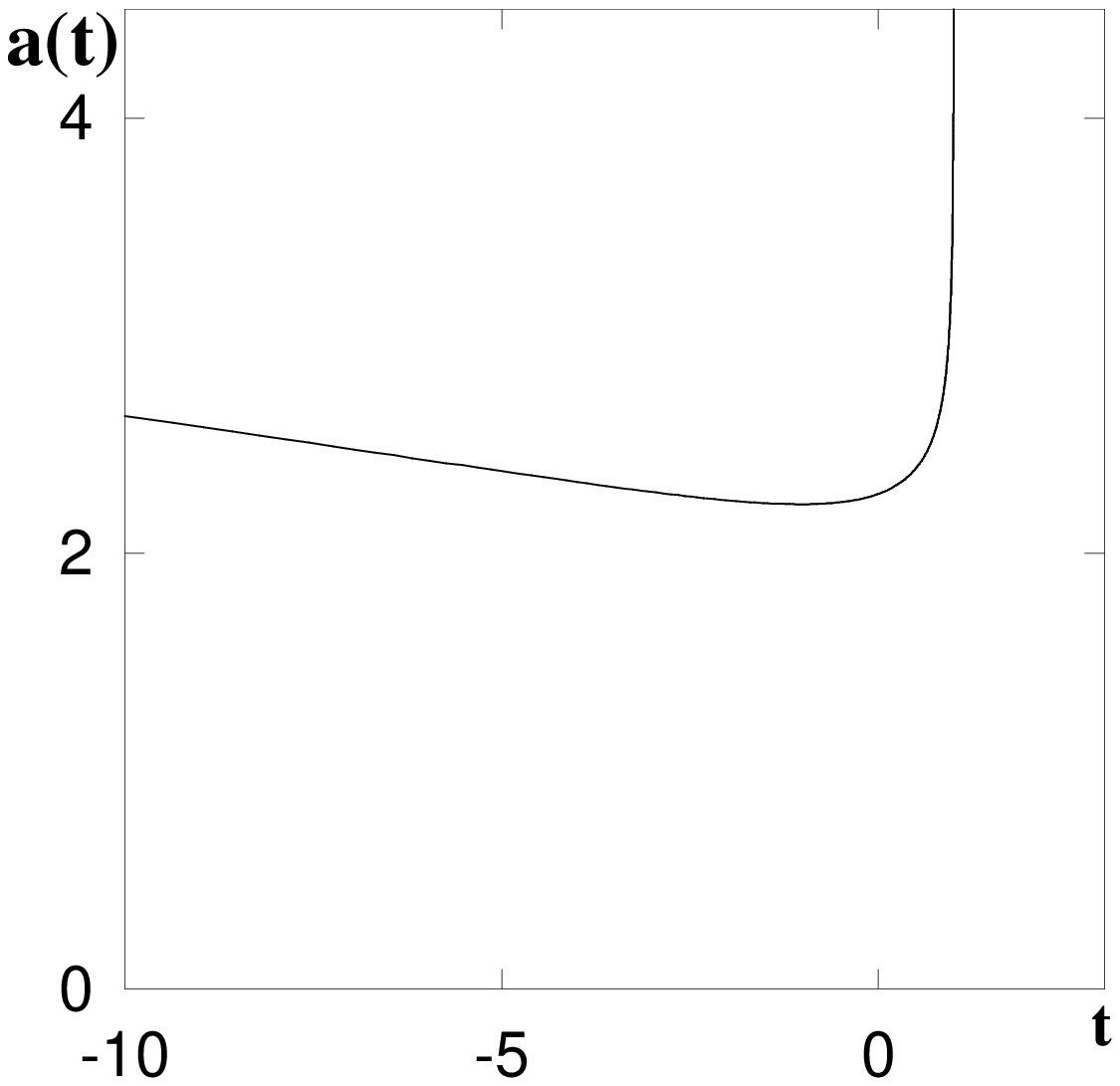,width=10cm,height=12cm}}
      %\put(30,30){\epsfig{file=phase31.ps,width=5cm,height=8cm}}
      %\put(70,30){\epsfig{file=phase32.ps,width=5cm,height=8cm}}
      %\put(-50,-30){\epsfig{file=phase4.ps,width=5cm,height=8cm}}
      %\put(-10,-30){\epsfig{file=phase5.ps,width=5cm,height=8cm}}
      %\put(30,-30){\epsfig{file=phase6.ps,width=5cm,height=8cm}}
      %\put(-35,40){$(a)$ phase $I$}
      %\put(5,40){$(b)$ phase $II$}
      %\put(45,40){$(c)$ phase $III^-$}
      %\put(85,40){$(d)$ phase $III^+$}
      %\put(-35,-15){$(e)$ phase $IV$}
      %\put(5,-15){$(f)$ phase $V$}
      %\put(45,-15){$(g)$ phase $VI$}
      \end{picture}
   \end{center}
%\caption{
\vspace{-3cm}
Fig. 1. The behavior of the scale factor with the D-particle.
%}
\label{fig1}
\end{figure}

\newpage
\vspace{1cm}

\begin{figure}
  \unitlength 1mm
   \begin{center}
      \begin{picture}(70,100)
      \put(-50,10){\epsfig{file=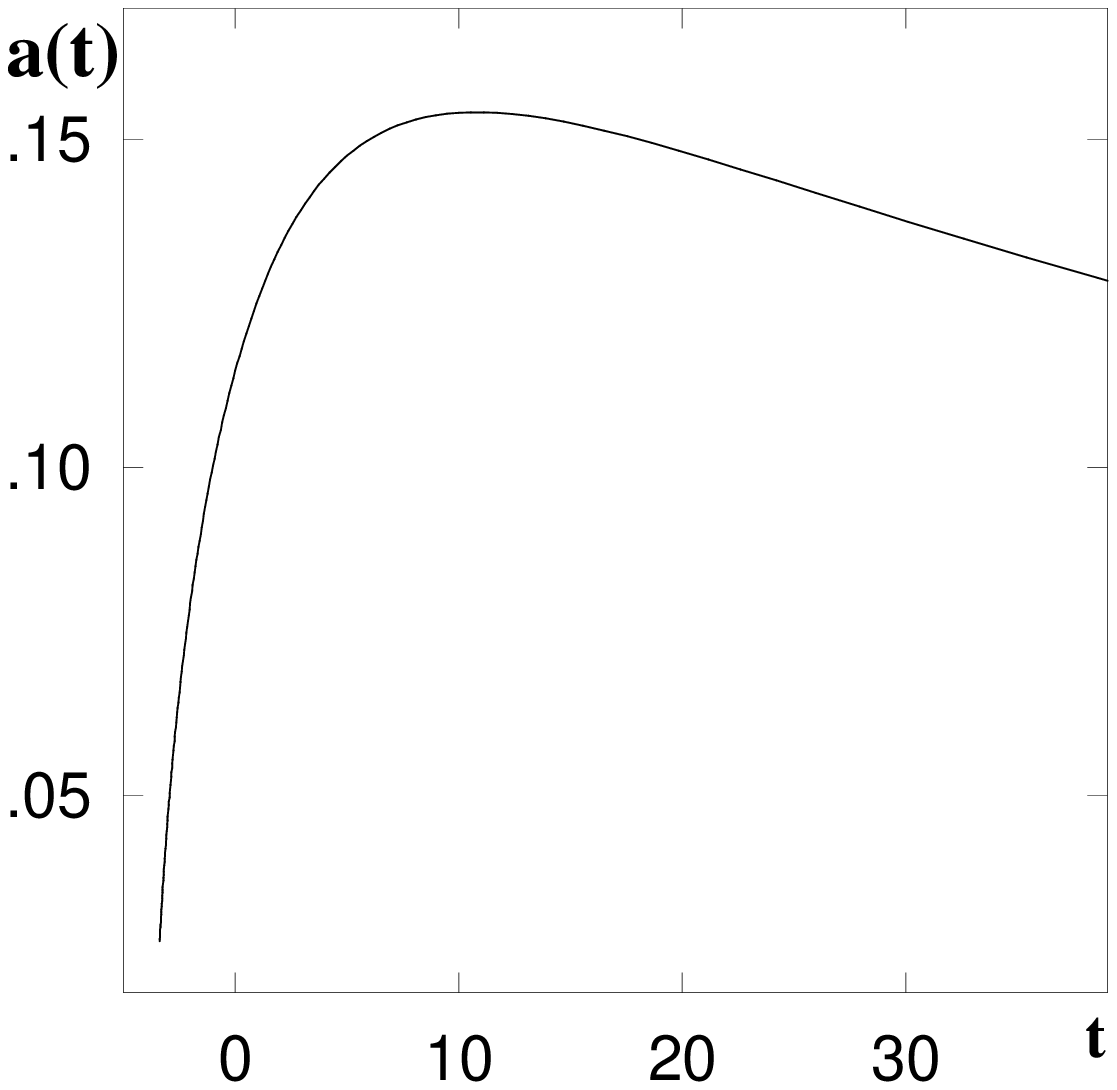,width=10cm,height=12cm}}
      \put(30,10){\epsfig{file=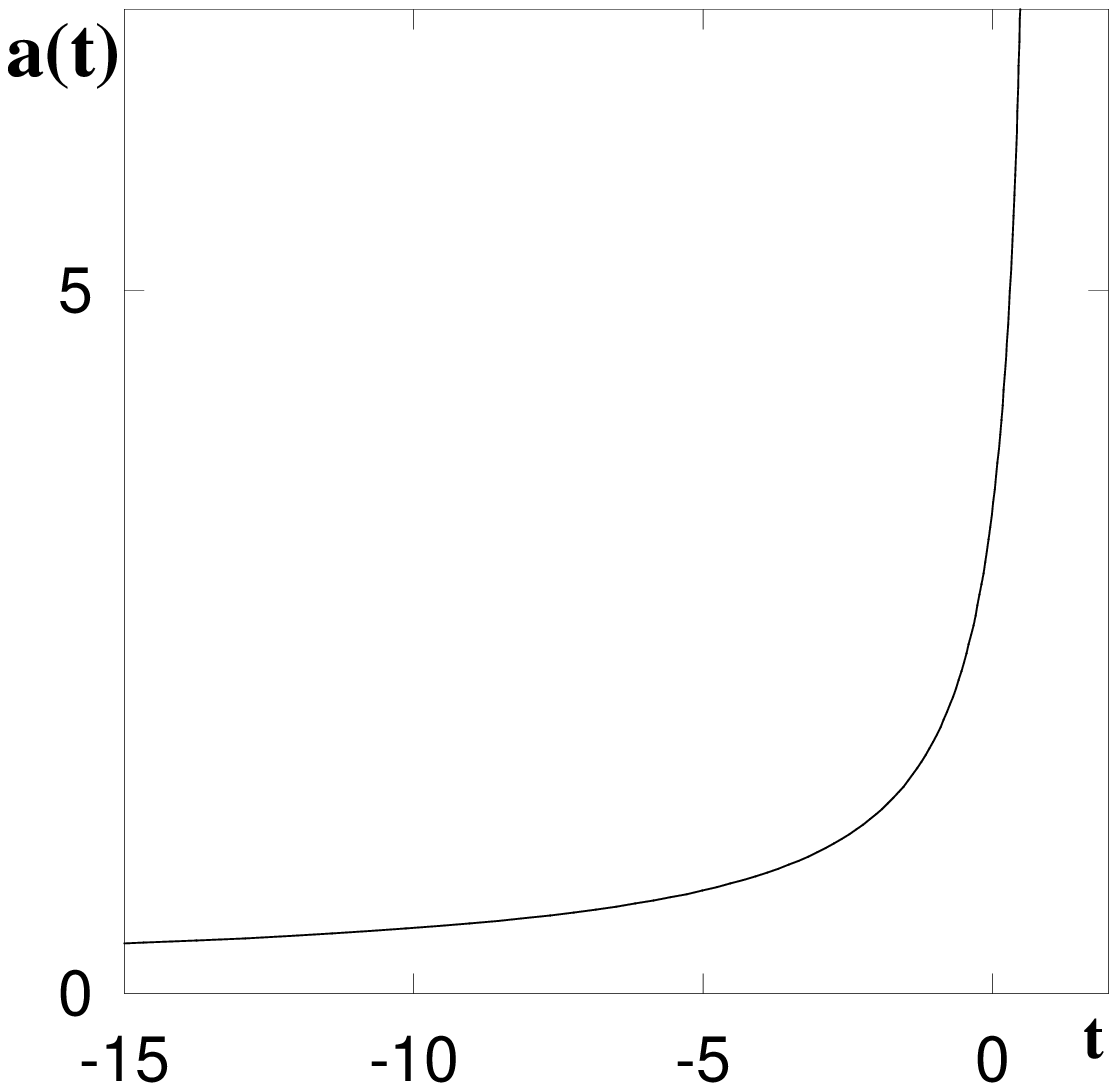,width=10cm,height=12cm}}
      \end{picture}
   \end{center}
%\caption{
\vspace{-4cm}
Fig. 2. The behavior of the scale factor with the D-string.
%}
\label{fig2}
\end{figure}

\newpage
\begin{figure}
  \unitlength 1mm
   \begin{center}
      \begin{picture}(70,100)
      \put(-10,10){\epsfig{file=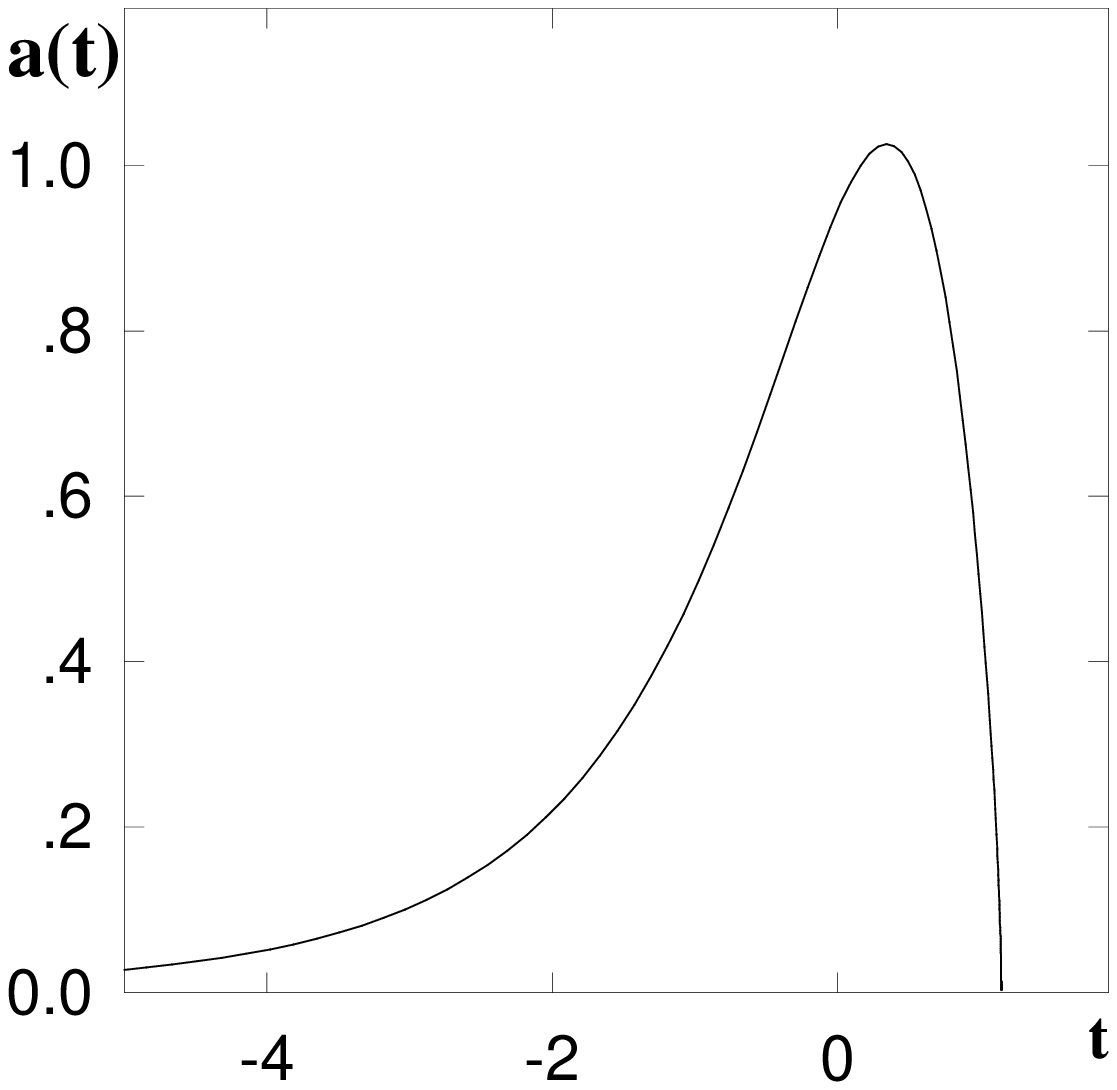,width=10cm,height=12cm}}
      \end{picture}
   \end{center}
%\caption{
\vspace{-4cm}
Fig. 3. The behavior of the scale factor with the Dp-brane ($p=2$ or $p=3$).
%}

\label{fig1}
\end{figure}

\end{document}